\documentclass[english,prc,aps,twocolumn,groupedaddress,nofootinbib,superscriptaddress,longbibliography,floatfix]{revtex4-1}
\usepackage{amsmath,amssymb,bm,graphicx,booktabs,array}
\usepackage[colorlinks=true,linkcolor=blue,citecolor=blue,urlcolor=blue]{hyperref}

\begin{document}

\title{Momentum Distributions and Spatial Signatures of Proton Halos in the {\it sd}
Shell}
\author{Taslima S.C. Diba}
\email{tdiba@leomail.tamuc.edu}
\affiliation{Department of Physics, East Texas A\&M University, Commerce, Texas 75429, USA}
\author{Carlos A. Bertulani}
\email{carlos.bertulani@etamu.edu}
\affiliation{Department of Physics, East Texas A\&M University, Commerce, Texas 75429, USA}
\affiliation{Technische Universit\"at Darmstadt, Institut f\"ur Kernphysik, 64289 Darmstadt, Germany}
\author{Ronaldo V. Lobato}
\affiliation{Centro Brasileiro de Pesquisas F\'isicas, Rua Dr. Xavier Sigaud, 150 - Urca, Rio de Janeiro, Brazil}
\email{lobato@cbpf.br}

%%%%%%%%%%%%%%%%%%%%%%%%%%%%%%%%%%%%%%%%%%%%%%%%%%%%%%%%%%%%%%%%%%%%%%%%%%%%%%%
\begin{abstract}
We perform a theoretical study of
intermediate-energy quasifree one-proton knockout reactions on proton
targets. Single-particle wave functions constrained by the experimental proton
separation energies are employed to calculate longitudinal momentum
distributions, one-proton removal cross sections, and full momentum-space
profiles for $^{26}$P,  $^{27}$S and $^{31}$Ar nuclei. To establish robust criteria to
identify proton halos, the analysis is extended beyond the traditional
momentum-width approach by investigating the spatial extension of the
valence proton through root-mean-square radii and the probability that the
proton resides outside the core nucleus,
$P(r>R_{\rm core})$. We also examine Coulomb-barrier
systematics, mirror-nucleus comparisons, realistic spectroscopic mixtures,
finite experimental momentum resolution, and uncertainties associated with
the proton separation energy.
Our calculations indicate that proton-halo structure cannot be identified reliably from a
single observable. A consistent interpretation emerges only when momentum distributions, spatial observables,
Coulomb effects, and many-body structure are considered simultaneously. Within
the present model, $^{26}$P exhibits the strongest proton-halo signatures, while $^{27}$S retains
pronounced halo-like features despite its larger Coulomb barrier. The more strongly
confined $^{31}$Ar provides a useful comparison and illustrates the progressive suppression of
halo observables with increasing binding and core charge. 
\end{abstract}
\maketitle

\section{Introduction}

The structure of atomic nuclei at the limits of stability
continues to reveal phenomena that have no counterpart among
stable isotopes. As the binding energy of the outermost nucleons
approaches zero, the coupling between bound states and the
particle continuum becomes increasingly important, giving rise
to extended matter distributions, modified shell structure,
strong configuration mixing, and new collective modes.
Radioactive-ion beam facilities have made it possible to
investigate these exotic systems over a large range of masses,
transforming weakly bound nuclei into one of the principal
frontiers of contemporary nuclear structure physics \cite{Tanihata1985PLB,Tanihata1985PRL,HansenJonson1987,Tanihata1992,Zhukov1993,BertulaniCantoHussein1993,Riisager1994,Tanihata1995,HansenJensenJonson1995,Tanihata1995,Ozawa2001,BertulaniHammerVanKolck2002,Michel2002,Jensen2004,Jonson2004,Thoennessen2004,TypelBaur2005,Michel2009,Hammer2017,Calci2016,Platter2017}.

Halo nuclei are characterized by
one or more valence nucleons whose wave functions extend far
beyond the range of the remaining nuclear core, producing
matter radii that are substantially larger than those expected
from the conventional $A^{1/3}$ dependence of nuclear size.
Since the pioneering interaction-cross-section measurements of
Tanihata \emph{et al.} \cite{Tanihata1985PLB,Tanihata1985PRL} and the subsequent theoretical
interpretation by Hansen and Jonson \cite{HansenJonson1987}, neutron halos have become
one of the best-established examples of quantum many-body
systems governed by continuum degrees of freedom. Extensive
experimental and theoretical investigations have identified
numerous neutron-halo nuclei, ranging from the light
Borromean systems $^6$He and $^{11}$Li to heavier candidates
such as $^{31}$Ne, establishing halos as a general phenomenon
of weakly bound nuclei \cite{Kelley1995,Aumann2000,Nakamura2014}.

The existence of proton halos is considerably more subtle.
Although a weakly bound proton also develops an extended
radial wave function, the long-range Coulomb interaction
introduces an additional confining potential that suppresses
the asymptotic tail relative to the neutron case.
Consequently, proton halos are expected to be less spatially
extended and more sensitive to the interplay between nuclear
binding, Coulomb repulsion, shell structure, and many-body
configuration mixing. The identification of proton halos
therefore requires considerably more stringent criteria than
those employed for neutron halos, and no single observable can
be regarded as definitive \cite{Schwab1995,Korolev2018,Csoto1994,Baye1995,Bertulani2005PLB}. Accordingly, the term ``proton halo'' should be reserved for cases in which weak binding, low orbital angular momentum, spatial extension, and reaction observables form a mutually consistent pattern rather than being inferred from a small separation energy alone.

This situation has changed significantly in recent years.
Precision mass measurements of proton-rich nuclei at and
beyond the proton drip line, combined with mirror-energy
difference (MED) analyses and modern \emph{ab initio}
VS-IMSRG calculations, have identified
$^{26}$P,
$^{27}$S,
and
$^{31}$Ar
as particularly promising proton-halo candidates \cite{Yu2024,Li2023CPC,Li2023PRC}.
The observed MEDs are substantially smaller than expected
from conventional shell-model calculations and can be
explained by an increased spatial extension of the weakly
bound $2s_{1/2}$ proton orbital, which reduces its average
Coulomb interaction with the core.
The corresponding proton separation energies,
approximately
118~keV,
869~keV,
and
674~keV,
together with the predicted occupancies of the
$2s_{1/2}$ orbital,
place these nuclei among the most favorable proton-halo
candidates presently known in the \emph{sd} shell.  
These nuclei are selected because recent mirror-energy-difference analyses and many-body calculations indicate appreciable \(2s_{1/2}\) proton strength near threshold. They therefore permit a controlled comparison of systems with similar shell structure but different proton binding energies and Coulomb barriers.
The present work was motivated in part by the proposed R3B Early Science program aimed at measuring quasifree (p,2p) momentum distributions of the proton-rich nuclei 
$^{26}$P,  $^{27}$S, and  $^{31}$Ar \cite{R3BProposal2025}.

Transfer reactions provide an important complementary probe of weakly bound proton configurations. At energies near or below the Coulomb barrier, sufficiently peripheral proton-transfer reactions are dominated by the external part of the proton-core overlap and can therefore be used to determine its asymptotic normalization coefficient (ANC). For a bound charged particle, the ANC fixes the amplitude of the Whittaker-function tail and thus provides a direct measure of the asymptotic component relevant to halo formation. This approach has been applied, for example, to the proton-halo problem in $^{8}$B, where ANCs extracted from peripheral reactions were used to constrain the spatial extension of the valence proton \cite{Carstoiu2001,Keeley2007,Mukhamedzhanov2022}. Quasifree knockout at several hundred MeV/nucleon provides complementary information: the longitudinal residue-momentum distribution is sensitive to the orbital angular momentum and spatial extent of the removed proton, but the reaction need not be exclusively peripheral and can retain sensitivity to more than the extreme asymptotic tail. A combined program of transfer/ANC measurements and high-energy quasifree knockout would therefore provide particularly stringent constraints on proton-halo structure.

The most direct reaction observable sensitive to the spatial
extension of a weakly bound nucleon is the longitudinal
momentum distribution of the residual nucleus following
single-nucleon removal. According to the uncertainty
principle, an extended coordinate-space wave function
corresponds to a relatively narrow momentum distribution.
This relationship has been exploited successfully in neutron
halo studies for more than three decades and constitutes one
of the most robust experimental signatures of halo structure.
Recent advances in radioactive-ion beam facilities,
high-luminosity hydrogen targets,
large-acceptance magnetic spectrometers,
and precision tracking detectors have now made analogous
measurements feasible for proton-rich nuclei using
intermediate-energy quasifree
$(p,2p)$
reactions,
opening a new avenue for directly testing the proton-halo
hypothesis \cite{Sakurai2013,Kamigaito2012,Sakurai2018,Thoennessen2013,Nilsson2013,SPIRAL22007,SPIRAL22008,Cornell2005,Aumann2021Review,Balantekin2014,BigRIPS2008,Obertelli2014,TostevinGade2021,BertulaniDoornenbalObertelliUesaka2026}.

Although narrow momentum distributions have long been
recognized as one of the principal signatures of halo
structure, the interpretation of proton-removal observables
is considerably more involved than for neutron halos.
Finite Coulomb confinement, configuration mixing,
spectroscopic fragmentation, detector resolution, and
uncertainties in the proton separation energy may all modify
the measured  longitudinal momentum distributions and removal cross
sections. Consequently, a convincing identification of a
proton halo should not rely on a single experimental
observable, but instead emerge from the consistent set
of complementary observables that probe both the
momentum-space and coordinate-space properties of the
valence proton. This viewpoint provides the central
motivation for the present work.

The primary objective of this paper is to formulate quantitative and internally consistent indicators of proton-halo structure in \(^{26}\mathrm{P}\), \(^{27}\mathrm{S}\), and \(^{31}\mathrm{Ar}\).
To accomplish this goal, we employ the
Glauber-eikonal reaction developed by Aumann et al. \cite{AumannBertulaniRyckebusch2013}. The calculations provide
predictions for longitudinal momentum distributions,
one-proton removal cross sections,
and their corresponding full widths at half maximum
(FWHM), which constitute the primary observables in
intermediate-energy quasifree
$(p,2p)$
reactions.

To establish a direct connection between reaction
observables and the underlying spatial structure of the
valence proton, we also calculate the root-mean-square
(rms) radius of the single-particle wave function and the probability for the valence proton to reside outside an adopted geometric core radius.  In the present work, a convincing proton-halo assignment requires the simultaneous occurrence of: (i) weak proton binding, (ii) substantial low-\(\ell\), especially \(s\)-wave, spectroscopic strength, (iii) an enhanced single-particle rms radius and exterior probability, and (iv) a narrow longitudinal momentum component that remains identifiable after configuration mixing and finite experimental resolution are included. 
These quantities provide complementary
information on the radial extension of the weakly bound
orbital and make it possible to correlate the narrowing of
the momentum distribution with the spatial development of
the proton halo.

A second objective of this work is to investigate the extent
to which the predicted halo signatures remain robust when
realistic physical effects are taken into account.
For this purpose we examine the influence of the Coulomb
barrier by comparing nuclei with similar binding energies
but different core charges, perform mirror-nucleus
comparisons that isolate Coulomb effects from nuclear
structure, construct realistic inclusive momentum
distributions using shell-model and
VS-IMSRG spectroscopic strengths,
study the degradation of the intrinsic momentum
distributions by finite experimental momentum resolution
through explicit convolution of the calculated spectra, and
propagate the experimental uncertainties of the proton
separation energies into the predicted momentum
distributions and FWHM values.
Together these studies determine the degree to which the
expected proton-halo signatures survive under realistic
experimental conditions.

This paper is organized as follows. Section~II  introduces the Glauber--eikonal
description of quasifree $(p,2p)$ knockout reactions, the construction of the
single-particle proton wave functions, the treatment of spectroscopic
strengths, and the definitions of the momentum-space and coordinate-space
observables used to characterize proton-halo structure.
Section~III presents the principal results for the proton-halo candidates
$^{26}$P, $^{27}$S, and $^{31}$Ar and for the neighboring isotopic chains.
The calculated longitudinal momentum distributions are examined
together with the corresponding single-particle rms radii and exterior
probabilities. 
Section~IV goes beyond the pure single-particle description and examines
the robustness of the predicted proton-halo signatures under more realistic
nuclear-structure and experimental conditions. 
Finally, Sec.~V summarizes the principal conclusions of the work and
discusses their implications for future quasifree $(p,2p)$ measurements of
proton-rich nuclei. 

\section{Theoretical Framework}

The identification of proton halos through quasifree one-proton knockout
reactions requires a theoretical description that consistently combines the
reaction dynamics with realistic single-particle structure. Since the
observables considered in the present work are sensitive to both the spatial
extension of the valence proton and the reaction mechanism, all calculations
have been performed within a single theoretical framework. This approach
ensures that momentum-space observables, coordinate-space observables, and
their systematic variations are derived from identical nuclear wave functions,
optical potentials, and reaction dynamics, thereby allowing direct physical
comparisons among the different observables investigated throughout this
paper.

The calculations are based on the Glauber-eikonal reaction model for (p,2p) reactions developed in Ref. \cite{AumannBertulaniRyckebusch2013,BertulaniDeConti2010}, which has been extensively applied to intermediate-energy
nucleon-removal and quasifree scattering reactions \cite{Movsesyan2013,Panin2016,Atar2018,DiazFernandez2018,Panin2019,Holl2019,Liu2019,Syndikus2020,Panin2021,Bertulani2021PRCL,Lehr2022,LiBertulaniXu2022,Panin2026,BertulaniLobato2026}. At beam energies of a few
hundred MeV/nucleon the collision time is short compared with the
characteristic time scale of the intrinsic nuclear motion, making the sudden
approximation well justified. Within this regime the reaction factorizes into
a description of the elementary proton-proton collision together with the
survival probabilities of the incoming projectile, the outgoing nucleons, and
the residual nucleus. The reaction amplitudes include both elastic and
absorptive interactions through eikonal $S$ matrices, providing a realistic
description of attenuation effects while maintaining the computational
efficiency required for systematic calculations over a broad range of binding
energies and nuclear species  \cite{AumannBertulaniRyckebusch2013}.

\subsection{Quasifree knockout reaction model}

In the quasifree A(p,2p)B reaction, a fast projectile nucleus collides with a hydrogen target and one bound proton is removed through a  proton-proton collision. The residual system is treated predominantly as a spectator during the short collision time, so that the measured residue momentum distribution retains sensitivity to the bound-state overlap of the removed proton. The inclusive one-proton removal cross section is written as
\begin{equation}
\sigma_{-p}=\sum_{\alpha} C^{2}S_{\alpha}\,\sigma_{\mathrm{sp},\alpha},
\end{equation}
where \(C^{2}S_{\alpha}\) is the spectroscopic factor for removal from orbital \(\alpha\), and \(\sigma_{\mathrm{sp},\alpha}\) is the corresponding single-particle cross section.

The single-particle cross sections are calculated according to the formalism of Aumann \textit{et al.}~\cite{AumannBertulaniRyckebusch2013}. The imaginary part of the optical potential is constructed from free nucleon-nucleon cross sections fitted to experimental data and corrected for in-medium effects~\cite{BertulaniDeConti2010}. Proton and neutron density distributions are obtained from Skyrme-Hartree-Fock calculations with the SLy4 energy-density functional \cite{Chabanat1998}.
These densities determine the eikonal phase shifts and absorption probabilities entering the reaction amplitudes. For comparison,
selected calculations were also performed using droplet-model densities, allowing the sensitivity of the predicted observables to the assumed density
distributions to be assessed. The real part of the optical potential was obtained by folding the M3Y interaction with the same Hartree-Fock densities \cite{Bertsch1977,SatchlerLove1979}. 

The elementary proton-proton interaction, optical potentials, and eikonal
survival probabilities are treated consistently throughout all calculations \cite{AumannBertulaniRyckebusch2013,BertulaniDeConti2010}.
Unless otherwise stated, identical reaction parameters are employed for every
nucleus and every systematic study presented in this work. We use 450 MeV/u as representative of experiments with the energies available at the GSI-FAIR facility \cite{BertulaniDoornenbalObertelliUesaka2026}.

The longitudinal residue-momentum distribution is
\begin{equation}
\frac{d\sigma}{dp_{\parallel}}
=
\sum_{\alpha} C^{2}S_{\alpha}
\frac{d\sigma_{\mathrm{sp},\alpha}}{dp_{\parallel}}. \label{dparal}
\end{equation}
The sum is incoherent because distinct residue final states are populated.
Throughout the present work we calculate both pure single-particle
distributions and realistic inclusive distributions in order to investigate
the influence of configuration mixing on the predicted proton-halo
signatures. For fast nucleon-removal reactions, the longitudinal momentum
distribution of the residue generally provides a cleaner probe of the
intrinsic momentum content of the removed nucleon than the corresponding
transverse distribution \cite{BertulaniMcVoy1992,BertulaniHansen2004,Orr1992}.
In the eikonal regime, reaction-induced deflections act predominantly
in the transverse plane, so that transverse momentum distributions can
acquire additional contributions from diffraction, elastic scattering
of the residue, and Coulomb deflection. The longitudinal distribution
is generally less sensitive to these reaction-induced effects. This
behavior was demonstrated theoretically in Refs.\
\cite{BertulaniMcVoy1992,BertulaniHansen2004}, while the weak target
dependence of measured parallel momentum widths has provided
experimental support for their use as structural observables
\cite{Orr1992}. This relative advantage is not universal, however, and
depends on the reaction mechanism, beam energy, target, measured
fragment, and experimental acceptance.

\subsection{Single-particle structure model}

The bound-state wave functions were generated by solving the radial Schr\"odinger equation in a Woods-Saxon potential containing central, Coulomb, and spin-orbit terms. The radius and diffuseness parameters were fixed at \(r_0=1.2A^{1/3}\,\mathrm{fm}\) and \(a=0.6\,\mathrm{fm}\). These are standard single-particle choices used to provide a common geometry across the systematic study, while the depth is readjusted to the prescribed separation energy. Fixing the geometry is deliberate: it isolates binding-energy and Coulomb trends. The spin-orbit strength was \(-20\) MeV and the same geometry as the central part.  This procedure isolates the influence
of the binding energy on the asymptotic form of the wave function and
therefore permits systematic studies of proton halos under controlled
conditions \cite{AumannBertulaniRyckebusch2013}.

The weakly bound valence proton is assumed to occupy the
$2s_{1/2}$ orbital, whose absence of a centrifugal barrier makes it the
natural candidate for halo formation in the proton-rich
\emph{sd}-shell nuclei considered here. For the spectroscopic-mixture calculations, additional \(1d_{5/2}\) contributions are included using the spectroscopic strengths reported by the shell-model and VS-IMSRG studies cited in Refs.~\cite{Tsukiyama2011,Tsukiyama2012,Bogner2014,Li2023CPC,Li2023PRC}.
These mixtures provide realistic inclusive longitudinal momentum distributions that can be
compared directly with future experimental measurements.

\begin{table}[t]
\caption{
Structural input adopted for the separation-energy sensitivity calculations.
The effective binding energy is given by
$E_{\rm eff}=S_p+E_x$.
The quoted spectroscopic factors correspond to the partial occupancies used
to construct the inclusive momentum distributions.
The single-particle cross sections,
$\sigma_{\rm sp}$,
are obtained from the reaction model of Ref.  \cite{AumannBertulaniRyckebusch2013} for each effective binding energy
and are multiplied by the corresponding spectroscopic factors when
constructing the inclusive cross sections.
}
\label{tab:structure_input}
\centering
\begin{tabular}{cccccccc}
\hline\hline
Projectile &
$E_x$ &
$n\ell j$ &
$C^2S$ &
$\sigma_{\rm sp}$ \\
 &
(MeV) &
 &
 &
 &
(mb) \\
\hline

$^{26}$P &
0 &
$2s_{1/2}$ &
0.722 &
6.98 \\

$^{26}$P &
2 &
$1d_{5/2}$ &
5.20 &
5.80 \\

$^{27}$S &
0 &
$2s_{1/2}$ &
1.184 &
6.48 \\

$^{27}$S &
2 &
$1d_{5/2}$ &
5.50 &
5.61 \\

$^{31}$Ar &
0 &
$2s_{1/2}$ &
1.760 &
6.00 \\

$^{31}$Ar &
2 &
$1d_{5/2}$ &
5.60 &
5.22 \\

\hline
\end{tabular}
\end{table}

In Table \ref{tab:structure_input} we list  the spectroscopic factors used in this work and the weighted (p,2p) single-particle cross sections for projectiles at 450 MeV/u. The $C^2S$  values for the $2s_{1/2}$  ground states are taken from Ref. \cite{Li2023CPC}. For the $1d_{5/2}$  states we assume an excited energy  $E_x=2$ MeV with a partially occupied orbital, so that the summed removal strength should approximately satisfy the occupancy sum rule $\sum_\alpha C^2S_\alpha(n\ell j) \approx n_{n \ell j}$ . 
The VS-IMSRG calculations \cite{Li2023CPC}  show that the  $1d_{5/2}$ is  strongly, but not completely, occupied. Their calculations use a 
$^{16}$O core and the full $sd$ valence space. From that study we infer the values of $\sum_\alpha C^2S_\alpha(1d_{5/2})=5.2$, 5.5, and 5.6 for $^{26}$P, $^{27}$S, and $^{31}$Ar, respectively. These values are rough estimates setting the stage for our reaction studies.

\subsection{Calculated observables}

The principal experimental observables investigated in this work are the
longitudinal momentum distributions, one-proton removal cross sections, and
their corresponding full widths at half maximum (FWHM). Because the measured residue momentum distribution is modified by reaction dynamics and absorption, it is not a direct Fourier transform of the bound-state wave function. Nevertheless, its width retains a strong sensitivity to the spatial localization and orbital angular momentum of the removed proton.
Weakly bound and spatially extended orbitals therefore produce
characteristically narrow momentum distributions \cite{Tanihata1985PLB,Tanihata1985PRL}.

For a radial wave function normalized as \(\int_0^\infty |u(r)|^2\,dr=1\), the single-particle rms radius is
\begin{equation}
r_{\mathrm{rms}}=
\left[
\int_0^\infty r^2 |u(r)|^2\,dr
\right]^{1/2}.
\end{equation}
The exterior probability is defined as
\begin{equation}
P(r>R_{\mathrm{core}})
=
\int_{R_{\mathrm{core}}}^{\infty}|u(r)|^2\,dr.
\end{equation}
This quantifies the probability for the valence proton to reside outside an adopted geometric core radius. These quantities provide complementary measures of
the spatial extension of the valence proton and permit direct comparisons
between the calculated momentum widths and the underlying radial structure.  We adopt the value $R_{\rm core}=1.2(A-1)^{1/3}$ fm. 

The same reaction framework is subsequently employed to investigate Coulomb
systematics, mirror nuclei, spectroscopic-strength mixtures, detector
resolution, and uncertainties associated with the proton separation energy.
Because every observable is calculated using identical reaction dynamics and
identical single-particle wave functions, the various signatures of proton
halo formation may be compared on a common theoretical basis.
We now apply this framework to the three proton-halo candidates
$^{26}$P,
$^{27}$S,
and
$^{31}$Ar,
beginning with the momentum-space observables that constitute the primary
experimental signatures accessible through quasifree
$(p,2p)$
reactions.

\section{Results}

\subsection{Proton-halo candidates}

The nuclei $^{26}$P, $^{27}$S, and $^{31}$Ar were selected as representative
proton-rich systems in the \emph{sd} shell for which modern nuclear-structure
calculations predict significant occupancies of weakly bound
$2s_{1/2}$ proton orbitals \cite{Bogner2014,Yu2024,Li2023CPC,Li2023PRC}.
Their small proton separation energies place the relevant \(2s_{1/2}\) strength close to the particle-emission threshold, where the asymptotic wave function becomes sensitive to both nuclear binding and Coulomb confinement.
Among these nuclei,
$^{26}$P represents the most weakly bound case presently accessible,
with a proton separation energy close to
0.1~MeV.
The neighboring nuclei
$^{27}$S and
$^{31}$Ar possess somewhat larger separation energies but are predicted to
retain substantial $2s_{1/2}$ occupancies. The adopted proton ground states separation energies $S_p$ are  calculated using the AME2020 mass evaluation \cite{Wang2021AME}. 
Together these systems provide an excellent laboratory for investigating the
evolution of proton halos as both the binding energy and the Coulomb barrier
increase with atomic number.

In all calculations the proton separation energies were taken from the most
recent experimental mass evaluations whenever available, while the
spectroscopic factors employed in the inclusive calculations were obtained
from modern shell-model and VS-IMSRG calculations together with experimentally
motivated occupancy systematics.
For the present work, particular emphasis is placed on the comparison between
pure $2s_{1/2}$ single-particle calculations and realistic spectroscopic
mixtures including the dominant $1d_{5/2}$ contributions.
This comparison allows the extent to which the halo signature survives
configuration mixing to be evaluated quantitatively.

\subsection{Longitudinal momentum distributions}

\begin{figure}[t]
\centering
\includegraphics[width=\linewidth]{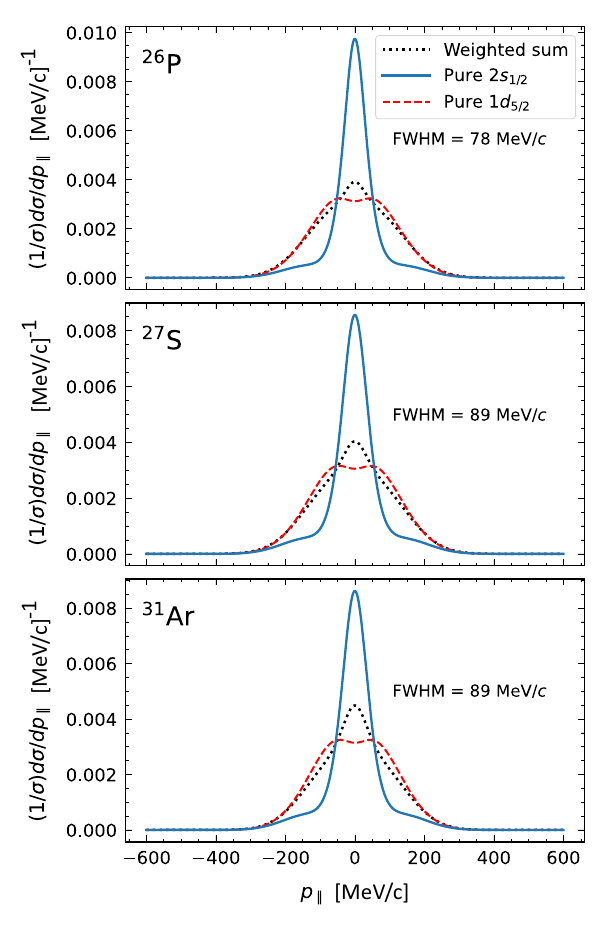}
\caption{Longitudinal momentum distributions for one-proton removal from \(^{26}\mathrm{P}\), \(^{27}\mathrm{S}\), and \(^{31}\mathrm{Ar}\) at 450 MeV/nucleon. The solid and dashed curves show unit-area-normalized \(2s_{1/2}\) and \(1d_{5/2}\) single-particle distributions, respectively, while the dotted curve is the spectroscopic-strength-weighted sum. Normalization to unit area, means $(d\sigma/dp_\parallel)/\sigma$, where $\sigma$ is the cross section for each case. The FWHM of the \(2s_{1/2}\)  distribution is indicated in each panel.  The FWHM of the \(1d_{5/2}\) and the mixed  distributions are given in the text. The adopted spectroscopic strengths are listed in Table~\ref{tab:structure_input}.}
\label{fig:momentum}
\end{figure}

Figure~\ref{fig:momentum} presents the calculated longitudinal momentum
distributions for proton removal from
$^{26}$P,
$^{27}$S,
and
$^{31}$Ar at a beam energy of
450~MeV/nucleon.
The calculations include the physically relevant proton orbitals,
allowing direct comparisons between weakly bound
$2s_{1/2}$ and
$1d_{5/2}$,
and mixed configurations.
Figure~1 compares the unit-area-normalized \(2s_{1/2}\) and \(1d_{5/2}\) components with their spectroscopic-strength-weighted sum. Normalization to unit area, defined as $(d\sigma/dp_\parallel)/\sigma$ where $\sigma$ is the cross section for each case, isolates differences in line shape, while the weighted sum illustrates how configuration mixing modifies the experimentally relevant inclusive profile. The FWHM values for the $2s_{1/2}$ orbitals extracted directly from the calculated line shapes are 78, 89, and 89 MeV/\(c\) for \(^{26}\)P, \(^{27}\)S, and \(^{31}\)Ar, respectively. Because the normalized distributions are non-Gaussian and exhibit different tails, their peak heights alone do not provide a reliable measure of their FWHM. For the $1d_{5/2}$ orbitals the corresponding widths are 290, 300, and 290 MeV/\(c\), and for the spectroscopically weighted distributions, the FWHM values are 250, 230, and 190 MeV/\(c\), respectively (\(^{26}\)P, \(^{27}\)S, and \(^{31}\)Ar).

Several systematic features emerge immediately.
For each of the three nuclei shown in Fig. \ref{fig:momentum}, the
$2s_{1/2}$ orbital produces the narrowest longitudinal momentum distribution,
consistent with the spatially extended wave function expected for a weakly
bound proton occupying an orbital without a centrifugal barrier.
The
$1d_{5/2}$  orbital exhibits a substantially broader distribution because the
centrifugal barrier localizes the wave function inside the nuclear surface.
The narrowing of the spectroscopically weighted momentum distributions therefore follows the
expected ordering and expected strength of orbital angular momentum and provides an immediate
signature of the underlying single-particle structure. The FWHM of the weighted sum longitudinal momentum distributions  decrease with the nuclei
 \(^{26}\)P, \(^{27}\)S, and \(^{31}\)Ar, in contrast to the widths of the pure $2s_{1/2}$ orbitals. This happens because the proportional contribution of the \(1d_{5/2}\) orbitals increases substantially for these nuclei, as shown in Table \ref{tab:structure_input}.

The calculations also demonstrate the strong influence of the proton
separation energy.
As the binding energy decreases, the asymptotic tail of the
$2s_{1/2}$ wave function extends progressively farther beyond the nuclear
surface, producing increasingly narrow momentum distributions.
This aspect is particularly pronounced for
$^{26}$P, whose extremely small proton separation energy results in the
narrowest calculated distribution among the three nuclei considered.
Although the Coulomb barrier suppresses the spatial extension relative to the
corresponding neutron case, the calculations clearly show that a narrow \(s\)-wave component remains present in the calculated spectrum and may be experimentally accessible if the resolution and statistical precision are sufficient.

\subsection{Spatial extension of the valence proton}

\begin{figure}[t]
\centering
\includegraphics[width=\linewidth]{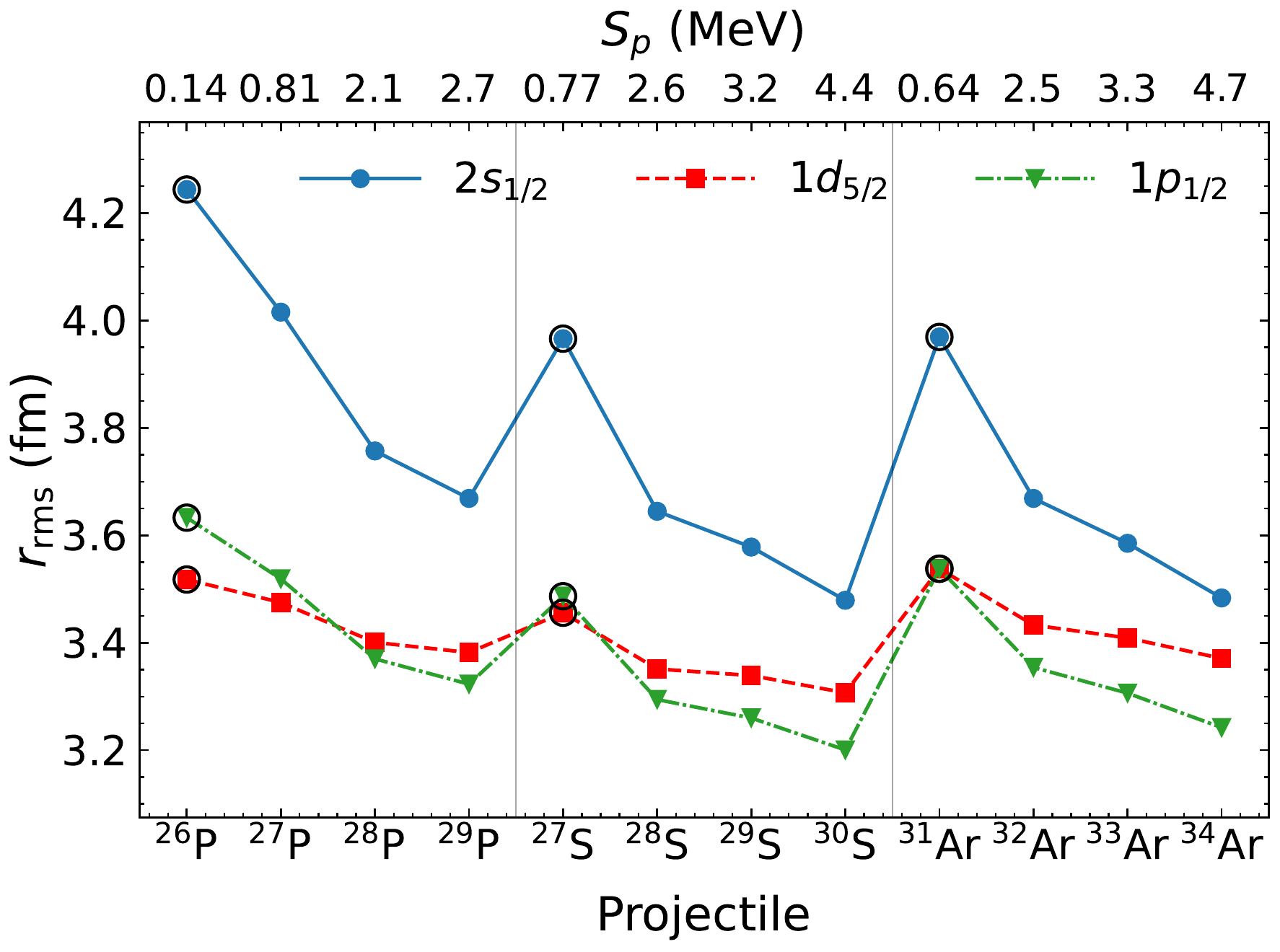}
\caption{
Calculated root-mean-square (rms) radii of the valence-proton
single-particle wave functions for proton removal from the proton-rich
isotopic chains $^{26-29}$P, $^{27-30}$S, and $^{31-34}$Ar. Results are shown
for the $2s_{1/2}$, $1d_{5/2}$, and $1p_{1/2}$ orbitals with the proton separation energies obtained from the 
AME2020 mass evaluation \cite{Wang2021AME}. The upper
horizontal axis displays the corresponding proton separation energies
$S_p$. Open circles identify the weakly bound nuclei $^{26}$P, $^{27}$S,
and $^{31}$Ar, which are the principal proton-halo candidates among the
nuclei considered. Vertical lines separate the phosphorus, sulfur, and
argon isotopic chains.
}
\label{fig:rms}
\end{figure}

The calculated rms radii for all investigated orbitals are summarized in
Fig.~\ref{fig:rms}. We include the 1$p_{1/2}$ orbital for a more intuitive comparison.
For every isotopic chain the weakly bound
$2s_{1/2}$ orbital possesses the largest rms radius, whereas the
$1d_{5/2}$ and
$1p_{1/2}$ orbitals remain considerably more localized.
The centrifugal barrier  suppresses the asymptotic penetration of the
$d$ wave, but it also repels the $d$-wave probability away from the nuclear
interior and concentrates it nearer the surface. By contrast, the
$1p_{1/2}$ wave function penetrates more strongly into the interior. Since
the rms radius weights the entire radial probability distribution by
$r^{2}$, not only its far asymptotic tail, the surface displacement of the
$1d_{5/2}$ density can give it a slightly larger rms radius than the
$1p_{1/2}$ state. The separately readjusted Woods--Saxon depths, spin-orbit
interaction, Coulomb field, and different shell structure further prevent
a universal monotonic ordering of rms radii based on $\ell$ alone.

The rms radius increases rapidly as the binding energy approaches the
particle-emission threshold, with
$^{26}$P exhibiting the largest calculated extension.
For a neutron \(s\)-wave, the asymptotic decay is approximately \(u(r)\propto e^{-\kappa r}\), with \(\kappa=\sqrt{2\mu S}/\hbar\). For a proton, however, the asymptotic form is a Coulomb Whittaker function. The exponential expression is therefore used only to illustrate the qualitative binding-energy dependence, while the actual calculations include the Coulomb potential explicitly.
A smaller proton separation energy therefore produces a slower asymptotic decay,
resulting in an increasingly diffuse proton density.
Although the Coulomb interaction suppresses this divergence relative to the
corresponding neutron case, the calculations clearly demonstrate that
substantial spatial extensions remain possible for weakly bound proton
orbitals.

\begin{figure}[t]
\centering
\includegraphics[width=\linewidth]{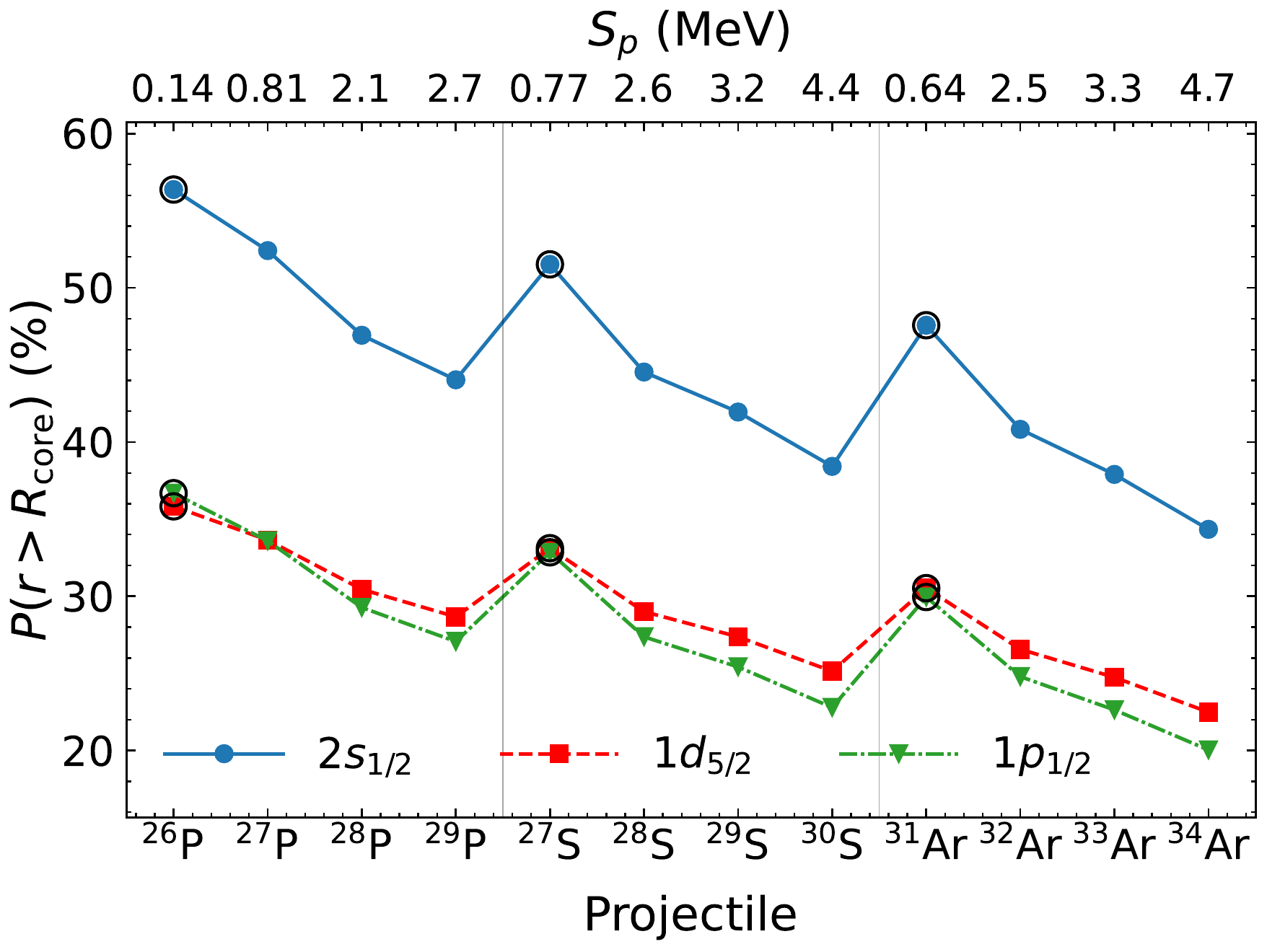}
\caption{
Calculated the probability for the valence proton to reside outside an adopted geometric core radius,  $P(r>R_{\rm core})$, for proton removal from the
proton-rich isotopic chains $^{26-29}$P, $^{27-30}$S, and $^{31-34}$Ar.  The upper horizontal axis displays the corresponding proton
separation energies obtained from the AME2020 mass evaluation \cite{Wang2021AME}. Open circles
indicate the proton-halo candidates $^{26}$P, $^{27}$S, and $^{31}$Ar.
}
\label{fig:p-outside}
\end{figure}

The calculated rms radii are fully consistent with the momentum
distributions presented in the previous section.
The nuclei possessing the largest rms radii also exhibit the narrowest
longitudinal momentum widths, confirming the expected inverse relationship
between coordinate-space localization and momentum-space extension.
The agreement between these independent observables provides strong evidence
that the calculated narrow momentum distributions originate from genuine
halo-like wave functions rather than from reaction-model artifacts.

The calculated external probabilities are shown in
Fig.~\ref{fig:p-outside}. For consistency, a single prescription for \(R_{\mathrm{core}}\) should be used throughout. We adopt \(R_{\mathrm{core}}=1.2(A-1)^{1/3}\) fm for all nuclei. The sensitivity of \(P(r>R_{\mathrm{core}})\) to reasonable variations of this radius is discussed below.
As expected, the
$2s_{1/2}$ orbitals exhibit  larger values than the
$d$-wave configurations for every isotopic chain.
The weakly bound
$^{26}$P calculation predicts the largest probability of the probability for the valence proton to reside outside an adopted geometric core radius, followed by
$^{27}$S and
$^{31}$Ar.
The ordering is identical to that observed for the rms radii and momentum
widths, demonstrating the strong correlation among these independent halo
observables.

The physical interpretation is straightforward.
For deeply bound orbitals the radial wave function decreases rapidly outside
the nuclear surface, producing a substantially smaller exterior probability than the weakly bound $2s_{1/2}$ configurations.
In contrast, weakly bound
$2s_{1/2}$ states possess long exponential tails extending over several
femtometers beyond the core.
The resulting increase of
$P(r>R_{\rm core})$
constitutes a direct manifestation of the spatial decoupling of the valence
proton from the remaining nucleons.
Unlike the rms radius, which receives contributions from the entire radial
wave function, the external probability isolates the asymptotic region and
therefore is a particularly sensitive measure of halo formation.

The close correlation between
$r_{\rm rms}$,
$P(r>R_{\rm core})$,
and the longitudinal momentum width demonstrates that these observables
provide complementary views of the same physical phenomenon.
The narrow momentum distributions originate from precisely those wave
functions possessing the largest external probabilities, illustrating the
fundamental connection between the asymptotic tail of the proton wave
function and the measured reaction observables.

\subsection{Systematics of longitudinal momentum widths}

\begin{figure}[t]
\centering
\includegraphics[width=\linewidth]{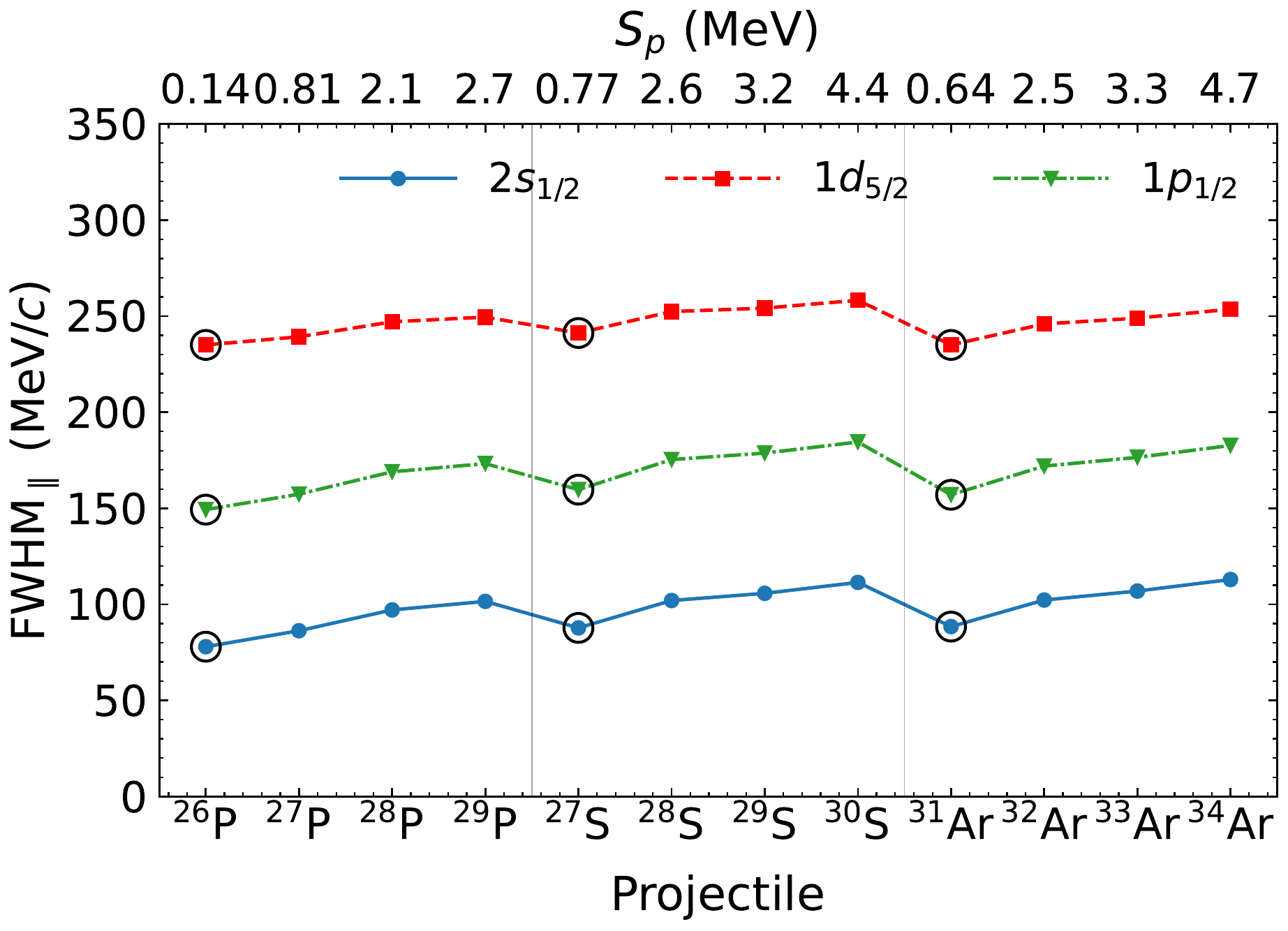}
\caption{
  Calculated full widths at half maximum (FWHM) of the longitudinal
  momentum distributions for one-proton removal from the proton-rich
  isotopic chains $^{26\text{--}29}$P, $^{27\text{--}30}$S, and
  $^{31\text{--}34}$Ar on a proton target at a beam energy of
  $450~\mathrm{MeV}/\text{nucleon}$. Results are shown for removal of a
  proton from the $2s_{1/2}$, $1d_{5/2}$, and $1p_{1/2}$ single-particle
  orbitals. 
  The nuclei $^{26}$P, $^{27}$S, and $^{31}$Ar, emphasized by open
  circles, are the principal proton-halo candidates in the three
  isotopic chains. The thin vertical lines separate the phosphorus,
  sulfur, and argon chains.
}
\label{fig:FWHM-chains}
\end{figure}

While the longitudinal momentum distributions contain the structural
information, their full widths at half maximum provide a convenient
single-parameter characterization that is particularly useful for systematic
comparisons between different nuclei and orbitals. Figure~\ref{fig:FWHM-chains} summarizes the calculated FWHM values for all proton
orbitals considered in the present work.
For every isotopic chain the
$2s_{1/2}$ orbital produces substantially smaller momentum widths than the
$d$-wave configurations.
This ordering reflects the inverse relation between spatial localization and momentum spread.

The systematic evolution of the FWHM also illustrates the role of the proton
separation energy.
Among the three nuclei,
$^{26}$P exhibits the smallest momentum width, followed by
$^{27}$S and
$^{31}$Ar.
This trend reflects the gradual increase of the effective confinement of the
valence proton produced jointly by the increasing proton separation energy and the
larger Coulomb barrier.
The calculations therefore demonstrate that the observed momentum width is
controlled not by the proton separation energy alone but by the combined influence
of the nuclear potential, the Coulomb interaction, and the orbital angular
momentum.

An important result of the present calculations is that the differences
between halo-like
$2s_{1/2}$ configurations and more localized
$d$-wave orbitals are considerably larger than the expected experimental
momentum resolution.
Consequently, the calculations predict that future high-resolution quasifree
knockout experiments should retain sensitivity to these configurations, provided that the experimental resolution, acceptance, and statistical uncertainties are smaller than the calculated separation between the corresponding line shapes.

\subsection{One-proton removal cross sections}

In addition to the longitudinal momentum distributions, the reaction calculations provide
predictions for the inclusive one-proton removal cross sections.
Within the spectroscopic-factor formalism the calculated cross section is
obtained as the sum of the single-particle contributions weighted by
their respective spectroscopic strengths.
The resulting observables therefore depend simultaneously on the reaction
dynamics and on the occupancies of the individual shell-model orbitals.

\begin{figure}[t]
\centering
\includegraphics[width=\linewidth]{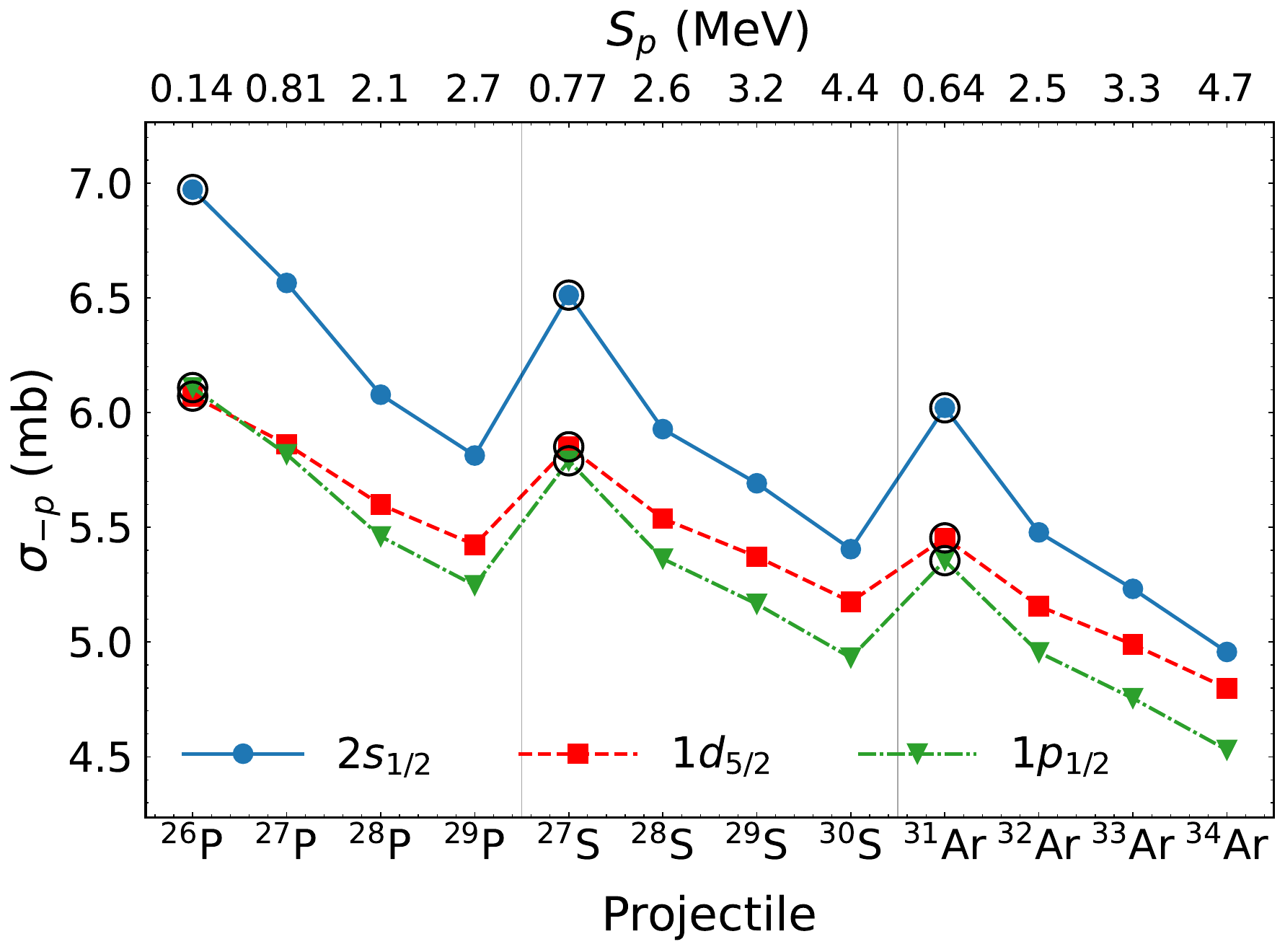}
\caption{
Calculated partial one-proton removal cross sections,
$\sigma_{-p}$, for proton-induced quasifree scattering reactions on the
proton-rich isotopic chains
$^{26-29}$P,
$^{27-30}$S,
and
$^{31-34}$Ar
at intermediate energies.
The results are shown separately for proton removal from the
$2s_{1/2}$,
$1d_{5/2}$,
and
$1p_{1/2}$
single-particle orbitals.
 Open circles identify the weakly bound nuclei
$^{26}$P,
$^{27}$S,
and
$^{31}$Ar,
which are among the most promising candidates for proton-halo structure.
}
\label{fig:cross}
\end{figure}

Figure~5 presents orbital-resolved one-proton removal cross sections for the \(2s_{1/2}\), \(1d_{5/2}\), and \(1p_{1/2}\) configurations. These quantities are single-particle reaction-model results and should not be interpreted as inclusive observables until they are multiplied by the corresponding spectroscopic strengths and summed over populated final states. The variation along each isotopic chain reflects changes in binding energy, radial overlap, absorption, and proton-core geometry.
The calculations indicate that the largest contributions originate from the
weakly bound
$2s_{1/2}$ and
$1d_{5/2}$ orbitals, reflecting both their large spectroscopic strengths and
their favorable single-particle removal probabilities.
Although the
$d$-wave configurations generally produce broader momentum distributions,
their larger occupancies contribute significantly to the inclusive removal
cross section and therefore cannot be neglected in realistic comparisons with
experiment.

The comparison between the longitudinal momentum distributions and the calculated cross
sections illustrates the complementary information carried by the two
observables.
Whereas the momentum width is primarily sensitive to the spatial extension of
the removed proton, the removal cross section depends more strongly on the
orbital occupancy and the reaction dynamics.
Only the simultaneous analysis of both observables therefore permits a
reliable determination of the underlying single-particle configuration.

Taken together, the calculated longitudinal momentum distributions and removal cross
sections establish clear theoretical benchmarks for future quasifree knockout
measurements.
The narrow momentum distributions predicted for weakly bound
$2s_{1/2}$ orbitals, together with the sizeable calculated removal cross
sections, demonstrate that the proposed proton-halo candidates appear favorable for experimental study. A quantitative feasibility statement would require the expected beam intensity, target thickness, acceptance, efficiency, and running time.

\subsection{Coulomb-barrier systematics}

\begin{figure}[t]
\centering
\includegraphics[width=\linewidth]{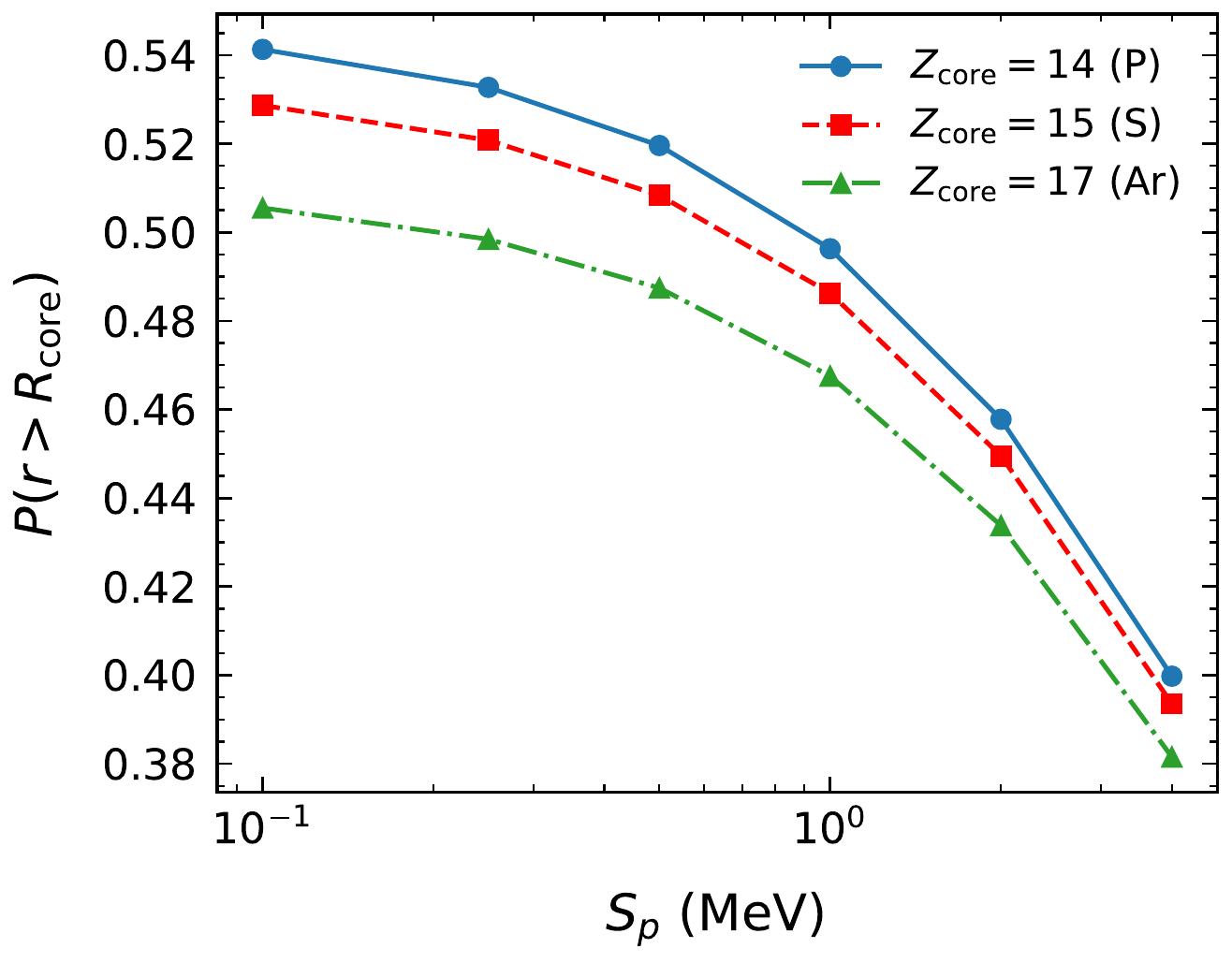}
\caption{
Calculated the probability for the valence proton to reside outside an adopted geometric core radius, $P(r>R_{\rm core})$, for hypothetical $2s_{1/2}$ proton states in the isobaric nuclei $^{30}$P, $^{30}$S, and $^{30}$Ar. The proton separation energies were constrained to identical prescribed values in order to isolate the effect of the Coulomb interaction.  Although all three systems possess identical proton separation energies, the exterior probability decreases systematically with increasing core charge, demonstrating the suppression of proton-halo formation by the Coulomb barrier.
}
\label{fig:Poutside_A30}
\end{figure}

An important question concerning proton halos is the extent to which the
Coulomb interaction limits the spatial extension of weakly bound proton
orbitals.
Unlike neutron halos, whose asymptotic tail is governed almost
exclusively by the proton separation energy, proton halos are simultaneously
influenced by the long-range Coulomb potential.
To isolate this effect, additional calculations were performed for
hypothetical
$2s_{1/2}$ proton states having identical separation energies but different
core charges corresponding to the phosphorus, sulfur, and argon isotopic
chains. For each imposed value of \(S_p\), only the central Woods-Saxon depth was readjusted; the projectile mass, radius and diffuseness parameters, spin-orbit term, orbital quantum numbers, beam energy, and reaction inputs were kept fixed.

\begin{figure}[t]
\centering
\includegraphics[width=\linewidth]{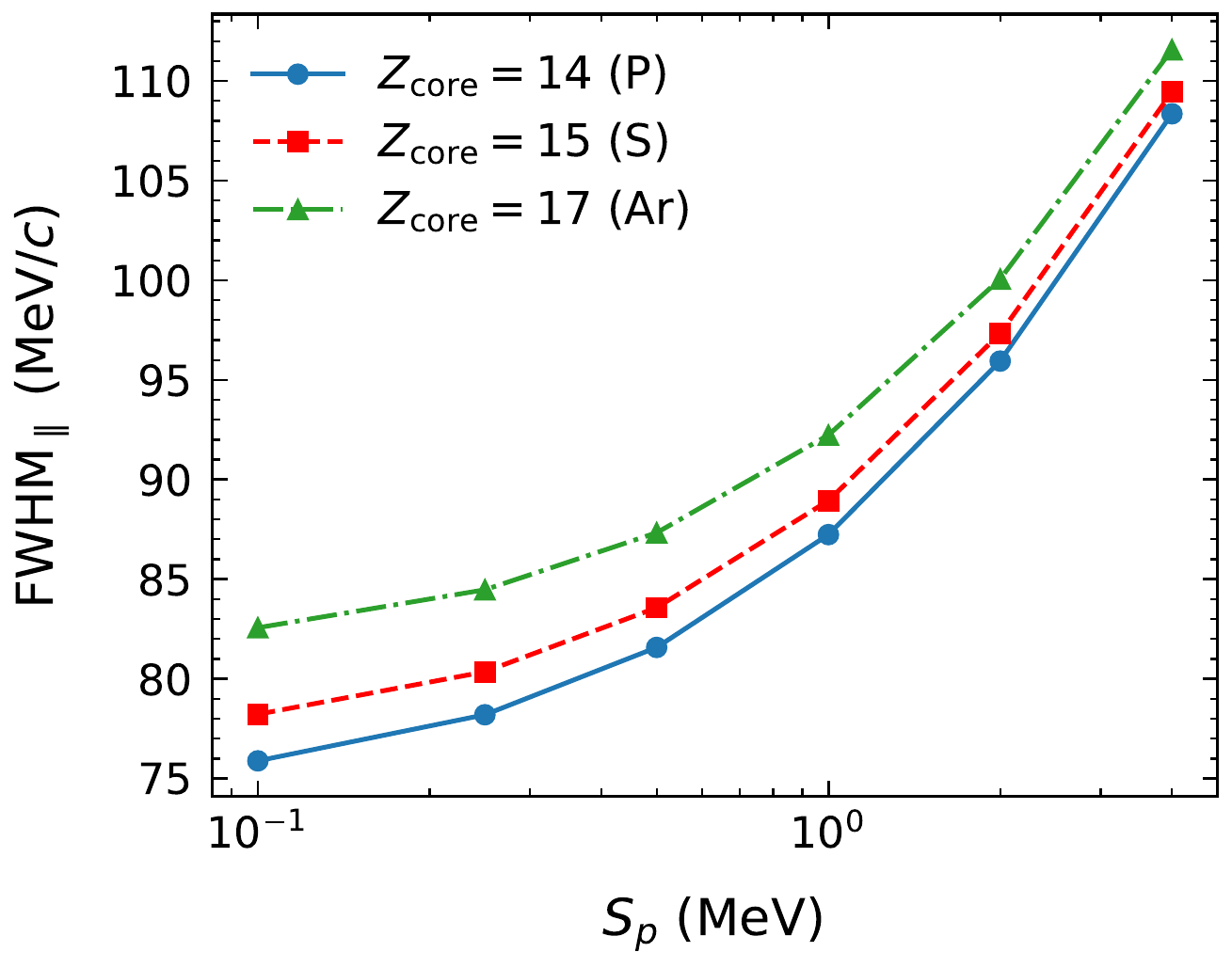}
\caption{
Calculated full widths at half maximum (FWHM) of the longitudinal momentum distributions for hypothetical $2s_{1/2}$ proton-removal reactions from the isobaric nuclei $^{30}$P, $^{30}$S, and $^{30}$Ar at a beam energy of 450 MeV/nucleon. For each nucleus the proton separation energy was artificially constrained to the same prescribed values, thereby isolating the influence of the Coulomb interaction on the asymptotic proton wave function. Since the projectile mass, Woods-Saxon geometry, orbital quantum numbers, reaction kinematics, and beam energy are identical in all calculations, the only significant structural difference among the three systems is the increasing core charge, $Z_{\rm core}=14$, 15, and 17. 
}
\label{fig:coulomb}
\end{figure}

The resulting external probabilities and longitudinal momentum distributions are presented
in Figs.~\ref{fig:Poutside_A30} and \ref{fig:coulomb}.
Because the proton separation energy is held fixed throughout the calculations, the
observed differences arise almost entirely from the increasing Coulomb
interaction.
The calculations show that the radial extension decreases systematically as
the core charge increases from phosphorus to argon.
Consequently, the corresponding longitudinal momentum distributions become
progressively broader despite the identical proton binding energy.

These calculations demonstrate that the proton separation energy alone is
insufficient to characterize proton halos.
Although weak binding remains the primary requirement for the formation of
an extended wave function, the Coulomb barrier introduces an additional
length scale that suppresses the asymptotic tail of the proton density.
The Coulomb interaction therefore shifts the onset of halo formation toward
smaller separation energies than would be expected for neutron-rich systems.

The calculated Coulomb systematics also provide a direct explanation for the
ordering of the momentum widths discussed previously.
The comparatively narrow momentum distributions predicted for
$^{26}$P result not only from its exceptionally small proton separation
energy but also from its smaller Coulomb barrier relative to sulfur and
argon.
Likewise,
$^{31}$Ar exhibits broader momentum distributions than would be expected
from its proton separation energy alone because the larger core charge confines the
valence proton more strongly.
The calculations therefore demonstrate quantitatively that proton halos are
controlled by the combined influence of weak binding and Coulomb
confinement.

More generally, the present results suggest that proton halos should
not be regarded simply as neutron halos shifted to the proton-rich side of
the nuclear chart.
Instead, they represent a distinct class of weakly bound quantum systems in
which the long-range Coulomb interaction modifies the asymptotic tail
of the wave function while preserving many of the characteristic features of
halo structure.
The calculated coordinate-space observables presented in this section
provide a direct physical explanation for the momentum-space signatures
discussed previously and demonstrate that the observed narrow longitudinal
momentum distributions originate from genuinely extended proton wave
functions.

\section{Beyond the Single-Particle Picture}

The calculations presented in Figs. \ref{fig:rms}-\ref{fig:coulomb}  were performed for
individual single-particle orbitals in order to establish the fundamental
relationship between the radial extension of the valence proton and the
corresponding reaction observables.
Although such calculations provide the clearest interpretation of the
underlying physics, real nuclei are not described by pure single-particle
configurations.
Instead, the nuclear wave function is distributed among several shell-model
orbitals through configuration mixing and many-body correlations.
Consequently, experimentally measured longitudinal momentum distributions correspond to
spectroscopic-strength-weighted sums over all populated final states rather
than to a single orbital.

An equally important question concerns the role of the Coulomb interaction.
Unlike neutron halos, whose asymptotic tail is governed almost
exclusively by the proton separation energy, proton halos experience the additional
long-range Coulomb potential.
Determining whether the experimentally observable signatures originate
primarily from weak binding or from Coulomb confinement therefore requires
comparisons that isolate these two effects.

\subsection{Spectroscopic-strength mixtures \label{specadmin}}

Experimentally observed quasifree knockout
cross sections correspond to the sum of contributions from all
occupied orbitals as in Eq. (\ref{dparal}), with the spectroscopic factors
$C_\alpha^2S$ associated with orbital $\alpha$.
In the calculations presented in Fig.  \ref{fig:momentum} the dominant
$2s_{1/2}$
and
$1d_{5/2}$
configurations were included using spectroscopic strengths derived from
modern shell-model calculations and constrained by experimental occupancy
systematics  \cite{Bogner2014,Li2023CPC,Li2023PRC}.
The resulting inclusive longitudinal momentum distributions therefore represent realistic
predictions for future quasifree knockout experiments.
The calculated  longitudinal momentum distributions are shown for
pure
$2s_{1/2}$,
pure
$1d_{5/2}$,
and the corresponding spectroscopic mixtures.
The pure
$2s_{1/2}$
configuration exhibits the narrow momentum distribution characteristic of a
weakly bound proton halo, whereas the
$d$-wave contribution is considerably broader because the centrifugal barrier
localizes the proton wave function inside the nuclear surface.

\begin{figure}[t]
\centering
\includegraphics[width=\linewidth]{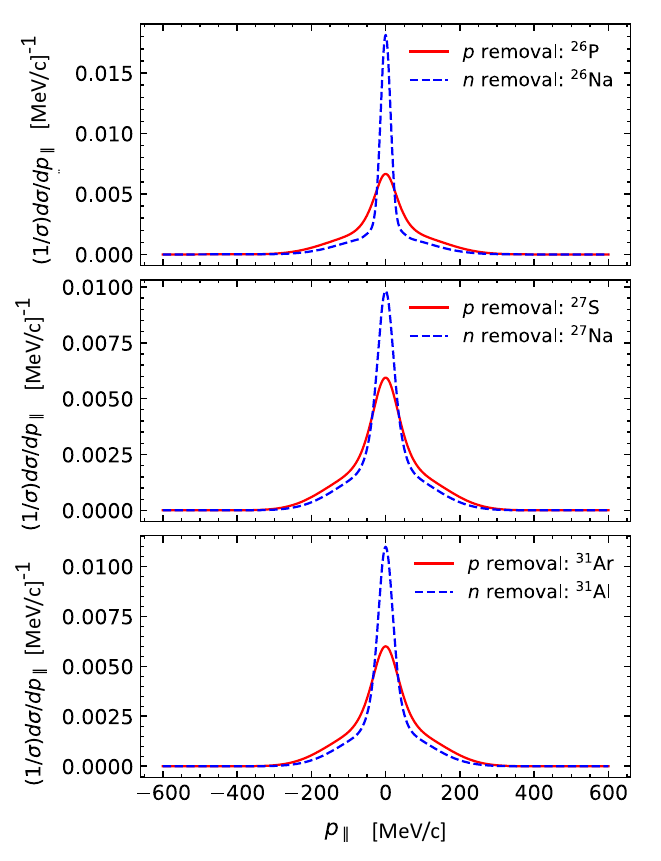}
\caption{
Calculated longitudinal momentum distributions for proton removal from
$^{26}$P, $^{27}$S, and $^{31}$Ar (solid curves) compared with neutron
removal from the corresponding mirror nuclei
$^{26}$Na, $^{27}$Na, and $^{31}$Al (dashed curves).
The calculations employ identical Woods-Saxon geometries, the same orbital
angular momentum, identical spectroscopic strengths, and matched effective
binding energies on the proton and neutron sides.
The only significant difference between each pair of calculations is therefore
the presence of the Coulomb interaction acting on the valence proton.
All distributions are normalized to unit area, $(d\sigma/dp_\parallel)/\sigma$,  to emphasize differences in
their shapes and widths rather than their absolute cross sections. 
}
\label{fig:mirror_distributions}
\end{figure}

The realistic inclusive distributions occupy an intermediate position.
Although the
$1d_{5/2}$
component increases the momentum width relative to the pure
$2s_{1/2}$
calculation, the narrow central peak produced by the weakly bound
$s$ wave remains clearly visible.
This result demonstrates that realistic shell-model configuration mixing does
not eliminate the principal halo signature.
Instead, the measured longitudinal momentum distribution retains clear
information regarding the underlying
$2s_{1/2}$
component even when substantial
$d$-wave occupancies are present.

\subsection{Mirror-nucleus comparisons}
Additional calculations were performed for the mirror pairs
$
^{26}\mathrm{P}
\leftrightarrow
^{26}\mathrm{Na},
$
$
^{27}\mathrm{S}
\leftrightarrow
^{27}\mathrm{Na}
$,
and
$
^{31}\mathrm{Ar}
\leftrightarrow
^{31}\mathrm{Al}.
$
The proton and neutron calculations employed identical Woods-Saxon
geometries, identical orbital angular momenta, identical spectroscopic
strengths, and matched effective binding energies.
The principal physical difference between the two calculations is therefore
the presence or absence of the Coulomb interaction. For each mirror pair, the proton and neutron calculations use the same orbital quantum numbers and Woods-Saxon geometry. The neutron potential depth is adjusted so that the neutron separation energy equals the effective proton binding energy used in the proton calculation. The effective binding energies (MeV) used for each of the pairs $^{26}$P/$^{26}$Na, $^{27}$S/$^{27}$Na, and $^{31}$Ar/$^{31}$Al  states are 0.118, 0.87, and 0.67,  for the 2$s_{1/2}$ states, and  $2.12=(0.12+2)$, $3.07=(0.87+2)$,  and $2.87=0.67+2$, for  the 1$d_{5/2}$, respectively. This artificial matching is intended to isolate the Coulomb contribution and is not a prediction for the physical neutron-rich mirror nucleus.

The calculated longitudinal momentum distributions are presented in
Fig.~\ref{fig:mirror_distributions}.
For every mirror pair the neutron-removal distribution is narrower than the
corresponding proton-removal distribution.
This behavior demonstrates directly that the Coulomb interaction suppresses
the radial extension of weakly bound proton orbitals even when the effective
binding energy is held fixed.

\begin{figure}[b]
\centering
\includegraphics[width=\linewidth]{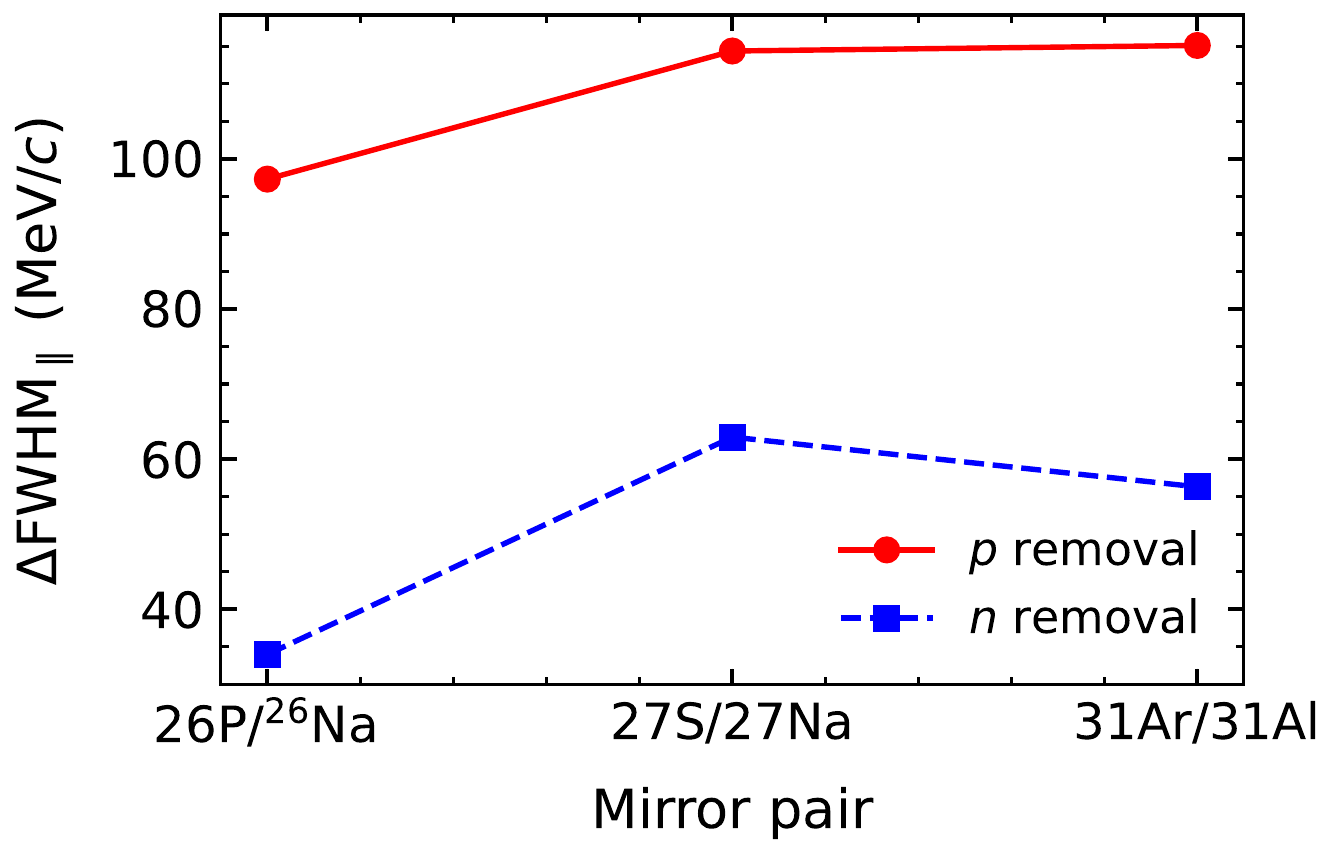}
\caption{
Calculated longitudinal momentum widths for proton removal
(red circles) and neutron removal (blue squares) for the three mirror
pairs considered in this work.
The calculations were performed using identical effective separation
energies and orbital occupancies, allowing the observed differences in
FWHM to be attributed primarily to Coulomb effects.
The differences $\Delta$FWHM for the proton-removal width from
$^{26}$P to $^{31}$Ar reflects the combined influence of increasing proton
binding and changing shell structure, whereas the neutron-removal widths
remain consistently smaller because of the absence of the Coulomb barrier.
}
\label{fig:mirror_fwhm}
\end{figure}

The corresponding momentum widths are summarized in
Fig.~\ref{fig:mirror_fwhm}.
To quantify the Coulomb effect we define
$
\Delta{\rm FWHM}_{\rm mirror}
$
which measures the difference between proton and neutron momentum widths for
otherwise equivalent mirror systems.
The calculated values of
$\Delta{\rm FWHM}_{\rm mirror}$
are positive for every nucleus investigated, confirming that proton removal
consistently produces broader momentum distributions than neutron removal.
This difference arises because the Coulomb barrier reduces the asymptotic
extension of the proton wave function while leaving the neutron wave function
unaffected.
The mirror calculations therefore isolate the influence of the Coulomb
interaction more clearly than direct comparisons between different
proton-rich nuclei.

An important consequence of these calculations is that proton halos should
not be viewed simply as neutron halos shifted to the proton-rich side of the
nuclear chart.
Instead, they constitute a distinct class of weakly bound quantum systems in
which the Coulomb interaction modifies the asymptotic wave function without
eliminating the fundamental halo phenomenon.
The quantity
$\Delta{\rm FWHM}_{\rm mirror}$
therefore emerges as a promising observable for future quasifree knockout
experiments designed to quantify the influence of Coulomb confinement on
weakly bound proton orbitals.

Finally, the mirror calculations provide an important consistency check for
the reaction model.
Because both members of each mirror pair are treated within the same
theoretical framework, many systematic uncertainties associated with the
reaction mechanism largely cancel in the difference.
Consequently, mirror comparisons offer one of the most direct theoretical
tests of proton-halo formation presently available and establish a close
connection between reaction observables and the underlying Coulomb
modification of the single-particle wave functions.

\section{Experimental Considerations}

Actual measurements are influenced by finite detector resolution,
acceptance effects, and experimental uncertainties in the proton separation
energies.
A realistic assessment of the feasibility of future quasifree knockout
experiments therefore requires an evaluation of the sensitivity of the
predicted observables to these effects.
We consider two important sources of uncertainty.
The first originates from the finite momentum resolution of the detector
system, which broadens the measured longitudinal momentum distributions.
The second arises from the experimental uncertainties of the proton
separation energies entering the bound-state calculations.
Together these studies suggest the robustness of the predicted halo
signatures under realistic experimental conditions.

\subsection{Influence of experimental momentum resolution}

The calculated  longitudinal momentum distributions  correspond to an
ideal detector possessing perfect momentum resolution.
In practice the measured distribution is the convolution of the intrinsic
reaction spectrum with the detector response function,
\begin{equation}
\left(\frac{d\sigma}{dp_{\parallel}}\right)_{\mathrm{obs}}
=
\int_{-\infty}^{\infty} dp_{\parallel}'\,
\left(\frac{d\sigma}{dp_{\parallel}'}\right)_{\mathrm{int}}
G(p_{\parallel}-p_{\parallel}';\sigma_M),
\end{equation}
where
\[
G(\Delta p;\sigma_M)
=
\frac{1}{\sqrt{2\pi}\sigma_M}
\exp\left[-\frac{\Delta p^2}{2\sigma_M^2}\right].
\]
is a normalized Gaussian having experimental momentum resolution
$\sigma_M$.
Unlike the commonly employed approximation based on quadratic addition of
widths, the present calculations perform the complete numerical convolution
of the calculated  longitudinal momentum distributions.
This procedure preserves the non-Gaussian features of the reaction spectra,
including the characteristic tails associated with different orbital angular
momenta.

\begin{figure}[t]
\centering
\includegraphics[width=\linewidth]{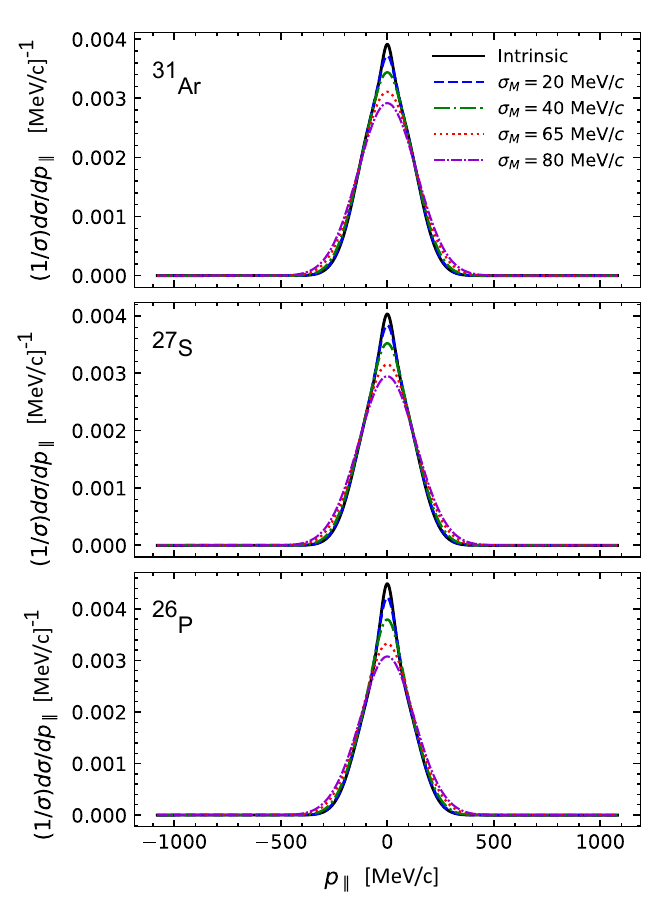}
\caption{Calculated longitudinal momentum distributions for one-proton removal from
$^{26}$P, $^{27}$S, and $^{31}$Ar at a beam energy of
450 MeV/nucleon after folding the intrinsic predictions with Gaussian experimental momentum resolutions.
The intrinsic distributions are shown by the solid curves, while the dashed, dash-dotted, dotted, and dot-dashed curves correspond to Gaussian missing-momentum resolutions of
$\sigma_M=20$, 40, 65, and 80 MeV/$c$, respectively.
The folded distributions were obtained by numerically convolving the complete calculated  longitudinal momentum distributions with the detector response function,
rather than by approximating the broadening through quadratic addition of widths.
All distributions are normalized to unit area, $(d\sigma/dp_\parallel)/\sigma$, in order to emphasize changes in their shapes and longitudinal widths.
}
\label{fig:resolution}
\end{figure}

To investigate the experimental sensitivity, the calculated momentum
distributions were folded using Gaussian resolutions of
$
\sigma_M
=
20,\;
40,\;
65,
\;\mathrm{and}\;
80~{\rm MeV}/c,
$
covering the range expected for present and forthcoming quasifree knockout
experiments.
The resulting folded  longitudinal momentum distributions are presented in
Fig.~\ref{fig:resolution}.
For the smallest resolutions,
20 and
40~MeV/$c$,
the intrinsic line shapes are essentially preserved.
The narrow
$2s_{1/2}$
component remains clearly distinguishable from the broader
$d$-wave contributions, indicating that halo signatures should be readily
observable with high-resolution detector systems.

As the detector resolution approaches
65--80~MeV/$c$,
the distributions become progressively broader and the differences between
the individual orbitals decrease.
Nevertheless, even under these conditions the weakly bound
$2s_{1/2}$
configurations retain noticeably narrower central peaks than the more deeply
bound
$d$-wave orbitals.
The calculations therefore demonstrate that finite detector resolution
reduces the experimental sensitivity but does not eliminate the principal
halo signatures.

\subsection{Sensitivity to proton separation energies}

\begin{figure}[t]
\centering
\includegraphics[width=\linewidth]{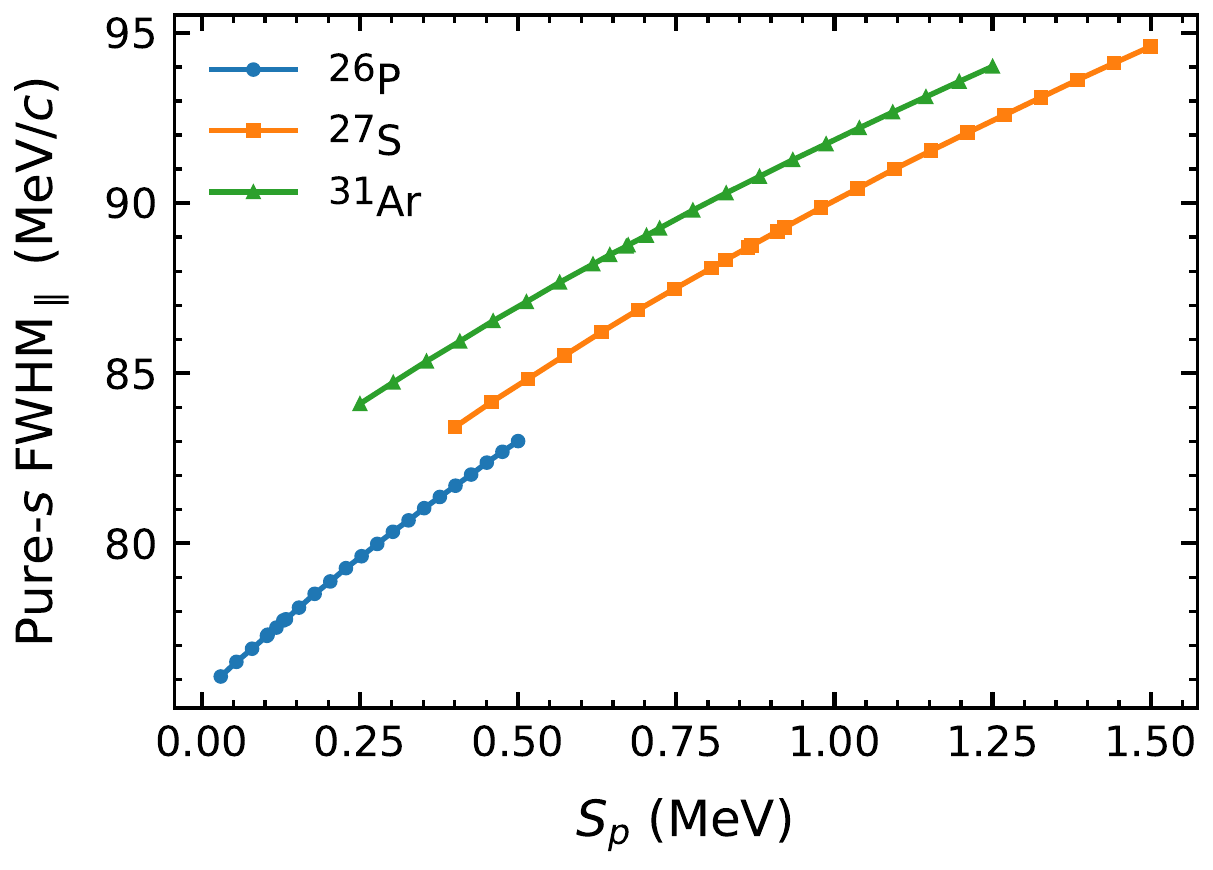}
\caption{
Calculated longitudinal momentum full widths at half maximum (FWHM) for
pure $2s_{1/2}$ proton removal from
$^{26}$P,
$^{27}$S, and
$^{31}$Ar as a function of the proton separation energy,
$S_p$.
For each isotope the Woods-Saxon potential depth was adjusted to reproduce
the prescribed proton separation energy while all remaining geometric parameters of
the single-particle potential were kept fixed.
The calculations were performed  at a beam
energy of 450~MeV/nucleon.
The figure illustrates the systematic evolution of the intrinsic momentum
width as the valence proton becomes progressively more tightly bound.
}
\label{fig:FWHM_vs_Sp}
\end{figure}

Figure \ref{fig:FWHM_vs_Sp} summarizes the intrinsic longitudinal momentum width
of a weakly bound proton as a continuous function of its separation energy.
For each nucleus, the depth of the Woods-Saxon binding potential was
adjusted so that the calculated $2s_{1/2}$ single-particle state reproduced
a prescribed value of the proton separation energy while the geometric
parameters of the potential remained unchanged.
The resulting wave functions were subsequently employed in the Glauber-based
reaction framework to calculate the longitudinal momentum
distributions following quasi-free one-proton knockout at
450~MeV/nucleon.
The FWHM values shown in the figure were extracted directly from the
complete calculated  longitudinal momentum distributions without assuming a Gaussian
functional form.
Consequently, the plotted widths represent genuine reaction-model
predictions rather than approximate estimates based on the second moments of
the distributions.

For all three nuclei the momentum width increases smoothly with increasing
proton separation energy, reflecting the progressive localization of the
single-particle wave function inside the nuclear potential.
This result follows directly from the relationship between
coordinate and momentum space.
As the proton becomes more deeply bound, the asymptotic exponential tail of
the wave function contracts, producing a broader momentum distribution.
Conversely, when the proton separation energy approaches the particle-emission
threshold, the wave function develops a long radial tail extending far
outside the nuclear surface, leading to increasingly narrow momentum
distributions.
The nearly linear behavior obtained over the investigated range indicates
that the reaction observable varies in a stable and predictable manner with
the proton binding energy, without exhibiting discontinuities or numerical
instabilities close to threshold.

The calculations also reveal systematic differences among the three isotopic
chains.
For a given proton separation energy the predicted FWHM values for
$^{26}$P are consistently smaller than those for
$^{27}$S and
$^{31}$Ar, whereas
$^{31}$Ar generally exhibits the broadest momentum distributions.
This ordering demonstrates that the proton separation energy alone does not
completely determine the momentum width.
The different core charges and masses modify the effective single-particle
potential through the Coulomb interaction and the reduced mass of the
proton-core system, producing modest but systematic shifts in the radial
extent of the valence orbital even when the binding energy is similar.
The figure therefore shows that proton-halo observables depend on the
combined influence of the proton separation energy and the underlying
mean-field potential rather than on either quantity independently.

\section{Conclusions and Outlook}

We have presented a comprehensive theoretical investigation of
proton-halo signatures in the proton-rich nuclei
$^{26}$P,
$^{27}$S,
and
$^{31}$Ar through quasifree one-proton knockout reactions at intermediate
energies.
The calculations were performed within the Glauber-based reaction 
model of Aumann et al \cite{AumannBertulaniRyckebusch2013}   using realistic single-particle wave functions constrained by
the experimental proton separation energies.
We studied momentum distributions, removal cross sections, radial observables, Coulomb-barrier systematics, mirror comparisons, spectroscopic mixtures, and detector-resolution effects within a common framework.

Within the adopted structure and reaction model, longitudinal momentum distributions are among the most sensitive observables considered here.
Weakly bound $2s_{1/2}$ proton orbitals produce narrow longitudinal momentum
distributions that are substantially different from those associated with
the more localized $1d_{5/2}$ configurations.
The predicted widths follow directly from the spatial extension of the
single-particle wave functions and therefore provide a sensitive probe of the spatial extension and momentum content of the weakly bound proton overlap.
The  calculations  demonstrate that
the nuclei possessing the narrowest momentum distributions also exhibit the
largest rms radii and the largest probabilities for the proton to reside
outside the nuclear core.

A second important conclusion concerns the role of the Coulomb interaction.
The additional calculations performed for hypothetical systems having
identical separation energies but different core charges demonstrate that
weak binding alone is insufficient to characterize proton halos.
The Coulomb barrier suppresses the radial extension of the valence proton,
broadening the corresponding  longitudinal momentum distributions even when the binding
energy is unchanged.
Mirror-nucleus calculations provide independent confirmation of this
aspect by showing that proton-removal  longitudinal momentum distributions are
systematically broader than neutron-removal distributions for otherwise
equivalent mirror systems.
These calculations quantify how Coulomb confinement modifies otherwise weakly bound proton orbitals and support treating proton-halo candidates separately from their neutron-halo analogues.

The spectroscopic-mixture calculations constitute another important result
of the present investigation.
Although realistic shell-model occupancies introduce significant
$1d_{5/2}$ contributions into the inclusive  longitudinal momentum distributions, the
characteristic narrow component associated with the weakly bound
$2s_{1/2}$ orbital remains clearly identifiable.
Consequently, the calculations predict that realistic many-body
configuration mixing does not destroy the principal halo signature.
Instead, the measured  longitudinal momentum distributions preserve sufficient
information to distinguish halo-like configurations from more localized
single-particle states.
This conclusion considerably strengthens the prospects for future
experimental identification of proton halos through inclusive quasifree
knockout measurements.

The detector-resolution and uncertainty studies further demonstrate the
robustness of the predicted observables.
Complete numerical convolutions of the calculated  longitudinal momentum distributions
show that detector resolutions of approximately
40~MeV/$c$ preserve nearly all of the intrinsic structural information,
whereas even resolutions approaching
80~MeV/$c$ retain differences between halo and non-halo
configurations.

The combined calculations identify
$^{26}$P as the strongest proton-halo candidate among the nuclei
investigated here, owing to its exceptionally small proton separation
energy and correspondingly large spatial extension.
The nucleus
$^{27}$S likewise exhibits pronounced halo characteristics, while
$^{31}$Ar displays weaker but still measurable signatures of an extended
$2s_{1/2}$ proton configuration.
Future quasifree knockout experiments capable of measuring high-resolution
longitudinal momentum distributions, together with inclusive and exclusive
one-proton removal cross sections, should therefore provide decisive tests
of the theoretical predictions presented in this work.

The methodology developed here is readily applicable to a much broader
range of proton-rich nuclei.
Future studies employing microscopic overlap functions derived from
VS-IMSRG, coupled-cluster, Gamow shell-model, or continuum shell-model
calculations, together with improved reaction descriptions, will further
reduce the theoretical uncertainties associated with the present approach.
Systematic measurements of isotopic and mirror chains extending beyond the
\emph{sd} shell will also provide valuable insight into the evolution of
proton halos with increasing Coulomb barrier and nuclear mass.

The next generation of radioactive-ion beam facilities will make it
possible to test these predictions with unprecedented precision.
The combination of high-resolution quasifree knockout reactions,
coincident $\gamma$-ray spectroscopy, improved nuclear-mass measurements,
and modern microscopic structure calculations promises to establish a new
generation of stringent tests of proton-halo formation throughout the
proton-rich side of the nuclear chart.

\begin{acknowledgments}
We acknowledge useful discussions with Tom Aumann and Alexandre Obertelli. This work was supported by the U.S. Department of Energy under Grant No. DE-SC0026074 and by the ExtreMe Matter Institute EMMI at the GSI Helmholtzzentrum f\"ur Schwerionenforschung. R.V.L. is supported by INCT-FNA (Instituto Nacional de Ci\^encia e Tecnologia, F\'\i sica Nuclear e Aplica\c c\~oes), research Project No. 464898/2014-5, and also thanks to CAPES/CNPq.
\end{acknowledgments}

\medskip

\begin{center}
{\bf DATA AVAILABILITY}
\end{center}

The data that support the findings of this article are not
publicly available upon publication because it is not technically feasible and/or the cost of preparing, depositing, and
hosting the data would be prohibitive within the terms of this
research project. The data are available from the authors upon
reasonable request.

\end{document}